\documentclass[11pt,english]{article}

\usepackage[T1]{fontenc}
\usepackage[latin9]{inputenc}
\usepackage[letterpaper]{geometry}
\usepackage{fancyhdr}
\usepackage{amsmath}
\usepackage{graphicx}

\usepackage{epsfig}
\usepackage{calligra}
\usepackage{subfigure}
\usepackage{float}
\usepackage[usenames,dvipsnames]{color}

\makeatletter
\usepackage{babel}
\makeatother

\usepackage{babel}
\begin{document}

\chead{UNCLASSIFIED/UNLIMITED}

\lhead{\includegraphics{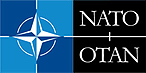}}

\rhead{\includegraphics[scale=0.1]{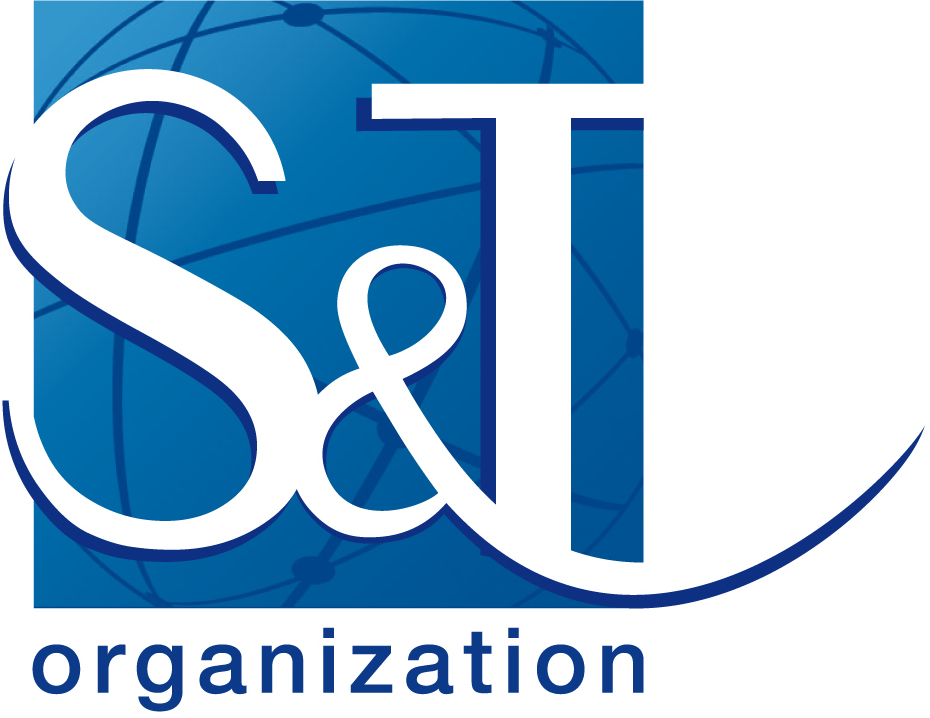}}

\cfoot{UNCLASSIFIED/UNLIMITED}
\title{\textbf{Interacting Swarm Sensing and Stabilization}}
\author{\textbf{Ira B. Schwartz$^{1}$, Victoria Edwards$^{2}$, and Jason
Hindes$^{3}$}\\
$^{1}$US Naval Research Laboratory, Code6792, Washington, DC 20375\\
 Email: ira.schwartz@nrl.navy.mil, Phone: 202 404 8359\\
 $^{2}$Mechanical Engineering and Applied Mechanics,\\
University of Pennsylvania,Philadelphia, PA 19104 USA\\
 Email:torriemed@gmail.com, Phone: 813 528 5240\\
 $^{3}$US Naval Research Laboratory, Code6792, Washington, DC 20375\\
 Email:jason.hindes@nrl.navy.mil, Phone: 810 434 0904}
\maketitle
\thispagestyle{fancy}

\begin{abstract}
Swarming behavior, where coherent motion emerges from the interactions of many mobile agents, is ubiquitous in physics and biology. Moreover, there are many efforts to replicate swarming dynamics in mobile robotic systems which take inspiration from natural swarms.  In particular, understanding how swarms come apart, change their behavior, and interact with other swarms is a research direction of special interest to the robotics and defense communities.   Here  we develop a theoretical approach that can be used to predict
the parameters under which colliding swarms form a stable milling state.
Our analytical methods rely on the assumption that, upon collision,
two swarms oscillate near a limit-cycle, where each swarm rotates
around the other while maintaining an approximately constant density.
Using our methods, we are able to predict the critical
swarm-swarm interaction coupling (below which two colliding swarms
merely scatter) for nearly aligned collisions as a function of physical 
swarm parameters. We show that the critical coupling corresponds to a saddle-node bifurcation
of a limit-cycle in the constant-density approximation. Finally, we show preliminary results from experiments in which two swarms of micro UAVs collide and form a milling state, which is in general agreement with our theory. 
 
\end{abstract}

\section{Introduction}
The emerging spatial-temporal motions of swarms of interacting agents
are a subject of great interest in application areas ranging from
biology to physics and robotics. Typically, swarming entails robust, self-organized motion, that 
emerges from the interaction of large numbers of simple mobile agents.
Examples have been observed in nature over many spatiotemporal scales
from colonies of bacteria, to swarms of insects\cite{Theraulaz2002,Topaz2012,Polezhaev,Li_Sayed_2012},
flocks of birds \cite{Leonard2013,Ballerini08,Cavagna2015}, schools
of fish\cite{Couzin2013,Calovi2014}, crowds of people\cite{Rio_Warren_2014},
and active-matter systems\cite{Cichos2020}. Understanding the underlying
physics behind swarming patterns and describing how they emerge from
simple models has been the subject of significant work in the mathematical
and engineering sciences \cite{Vicsek,Marchetti,Aldana,Desai01,Jadbabaie03,Tanner03b,Tanner03a,Gazi05,Tanner07}.
In pushing the theory to robotic platforms, engineers have focused
on designing and building swarms of mobile robots with a large and
ever expanding number of platforms, as well as virtual and physical
interaction mechanisms\cite{Cichos2020,AutonomousMobileRobots,8990018,7989200,MultiRobotSystems}.
Robotic applications range from exploration\cite{8990018}, mapping\cite{Ramachandran2018},
resource allocation \cite{Li17,Berman07,Hsieh2008}, and swarms for
defense \cite{Wong2020,Chung2011,Witkowski}

Since robotic swarms must operate in real environments, theoretical and experimental swarming systems have been analyzed in
many contexts, including swarms of mobile robots with homogeneous
and heterogeneous agents and delayed communication\cite{szwaykowska2016collective,edwards2020delay}.
Moreover, the dynamics of robotic swarms have been tested in complex
environments, from drones flying in the air, to boats tracking coherent
structures in complex flows, and collaborating robots locating sources
in turbulent media\cite{hajieghrary2016multi,heckman2015toward}. 

When deploying swarms in uncertain environments of varying complexity
and geometry, it is important to understand  stability.
Recently, we have analyzed stability of swarms in various configurations.
For example, we have studied swarms with complex network topology,
and quantified instabilities arising from heterogeneous topology in the number
of local interactions each agents has\cite{hindes2016hybrid}. We
have examined the effects of communication delay and how environmental
noise destabilizes self-organized patterns\cite{szwaykowska2018state,kyrychko2018enhancing}.
In addition, we have analyzed other environmental effects, such as
range-dependent communication and surface geometry, as a function
swarm control parameters \cite{hindes2020stability,hindes2020unstable}.

In all of the above--mentioned research we have considered only a
single swarm and its stability in complex environments. Here we extend
our analysis to multiple, interacting swarms, and their resulting
patterns. The general model that we use to describe the dynamics of both
single and interacting 
swarms contains self-propulsion, friction, and gradient-forces between agents:

\begin{equation}
\ddot{\bf{r}}_{i}= \big[\alpha_{i} -\beta|\dot{\bf{r}}_{i}|^{2}\big]\dot{\bf{r}}_{i}-\lambda_{i}\sum_{j\neq i} \partial_{\bf{r}_{i}}U(|\bf{r}_{j}-\bf{r}_{i}|)
\label{eq:swarmmodel}
\end{equation} 
where $\bf{r}_{i}$ is the position-vector for the $i$th agent
in two spatial dimensions, $\alpha_{i}$ is a self-propulsion constant,
$\beta$ is a damping constant, and $\lambda_{i}$ is a coupling constant\cite{Levine,Erdmann,DOrsagna,Romero2012}.
The total number of swarming agents is $N$, and each agent has unit
mass. Beyond providing a basis for theoretical insights, Eq.(\ref{eq:swarmmodel})
has been implemented in experiments with several robotics platforms
including autonomous ground, surface, and aerial
vehicles\cite{szwaykowska2016collective,edwards2020delay,hindes2020unstable}. We
remark that Eq.~(\ref{eq:swarmmodel})
contains most of the relevant physics needed to model an enormous class of
behaviors.Moreover,  additional physics, stochastic effects, and network
communication topologies may all be added to match many experiments.

In this paper, we restrict ourselves to  the well-known interaction Morse potential,
$U$, which controls local attraction and repulsion length scales
between interacting agents:
\begin{equation}
U(r)=Ce^{-r/l}-e^{-r}.\label{eq:MorsePotential}
\end{equation}

\section{The geometry and dynamics of colliding swarms}

\begin{figure}[h]
\begin{centering}
\includegraphics[scale=0.4]{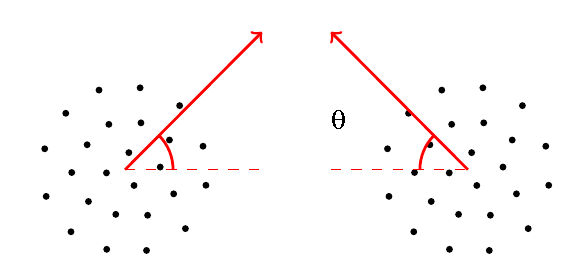}
\par\end{centering}
\caption{The geometry of two colliding swarms. The initial configurations are
flocking states, which intersect at an angle $\theta.$\label{fig:Geometry}}
\end{figure}

We use Eq.(\ref{eq:swarmmodel}) to model two interacting swarms with
the same underlying dynamics but different parameters and initial
conditions. The most straightforward collision scenario consists of two flocks
colliding, where each swarm has achieved velocity consensus well before
collision. The initial distance, $D$, which separates the swarms is large enough
so that the interaction forces between the swarms are exponentially
small, and $\theta$ defines the interaction angle. (See Fig.~\ref{fig:Geometry}.)
The potential function of the Morse potential is defined by

\noindent where $C,l$ define the repulsion and length constants respectively,
and the attraction length constant is scaled to unity.

\begin{figure}[h]
\begin{centering}
\includegraphics[scale=1.5]{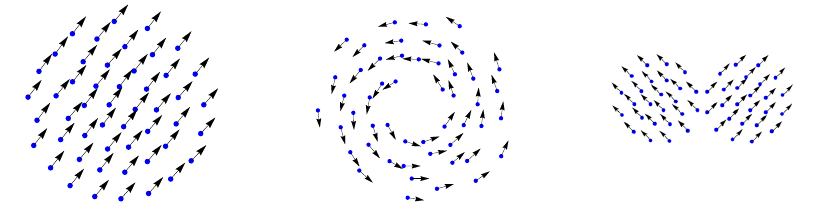}
\par\end{centering}
\caption{Possible final combined configurations of colliding swarm: flocking
states (left), milling states (middle), and scattering states (right).\label{fig:final-combined}}
\end{figure}

Given the initial flocking state configurations, there are three possible
final states of the combined interactions, shown in Fig.~\ref{fig:final-combined};
i.e., flocking where the swarms combined to form a translating state,
milling where the combined center of mass is stationary, or scattering
where the swarms pass through each other and flock in different directions.

\begin{figure}[t]
\begin{centering}
\includegraphics[scale=0.75]{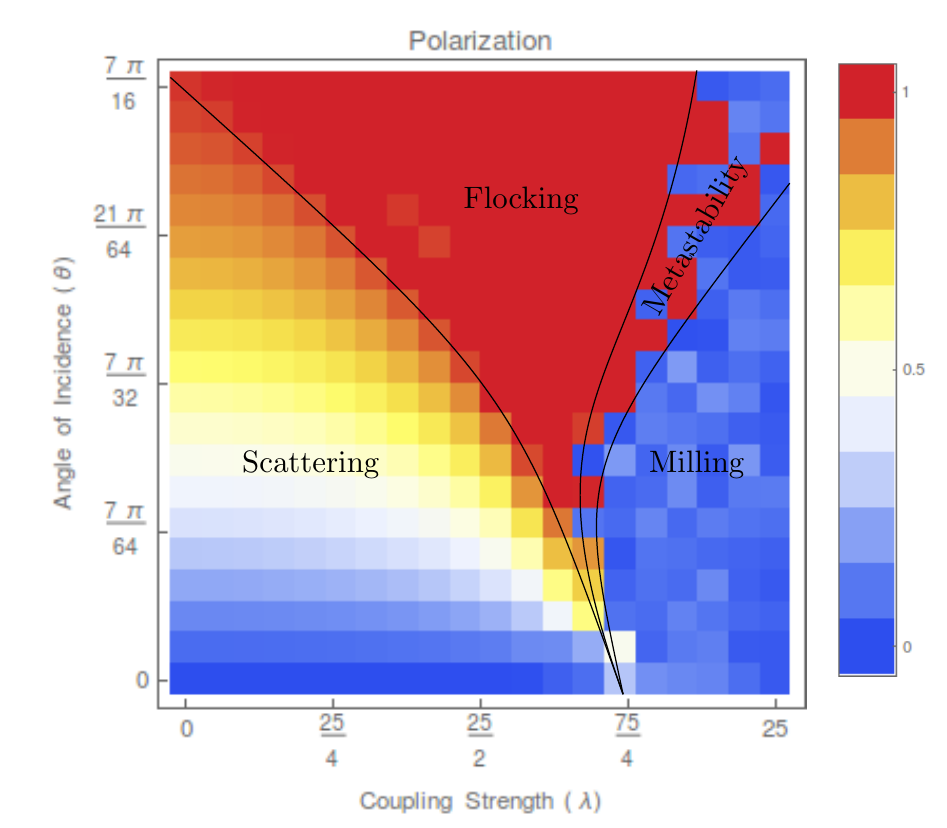}
\par\end{centering}
\label{fig:polariz}
\caption{Polarization as a function of collision
angle $\theta$ and coupling strength $\lambda$. See \cite{kolon2018dynamics} for details and parameter values.}
\end{figure}

A useful quantity for distinguishing between the three possibilities is the
polarization, $\cal P$, given by
\begin{equation}
{\cal P}=\frac{\lvert\sum_{i}
  \dot{\mathbf{r_i}}\rvert}{\sum_{i}\lvert\dot{\mathbf{r_i}}\rvert}.\
\label{eq:polarization}
\end{equation}
When the agents are in alignment, ${\cal P} \approx 1$, and it is
approximately zero when they are ani-parallel. Therefore,  when the swarms are in the
flocking state, ${\cal P}\approx 1$, while  in the milling state, ${\cal
  P}\approx 0$. When the swarm is  in the scattering state it is between 0
and 1. The polarization has been used to quantify the parameter space
comparing angle $\theta$ against the coupling strength $\lambda_i = \lambda$ in
Fig.~\ref{fig:polariz}. We notice that there exist distinct regions in
parameter space  where the milling state exists, as well as
other regions show the existence of scattering and flocking. 

\section{The milling state - stopping colliding swarms}

\begin{figure}[h]
\begin{centering}
\includegraphics[scale=0.25]{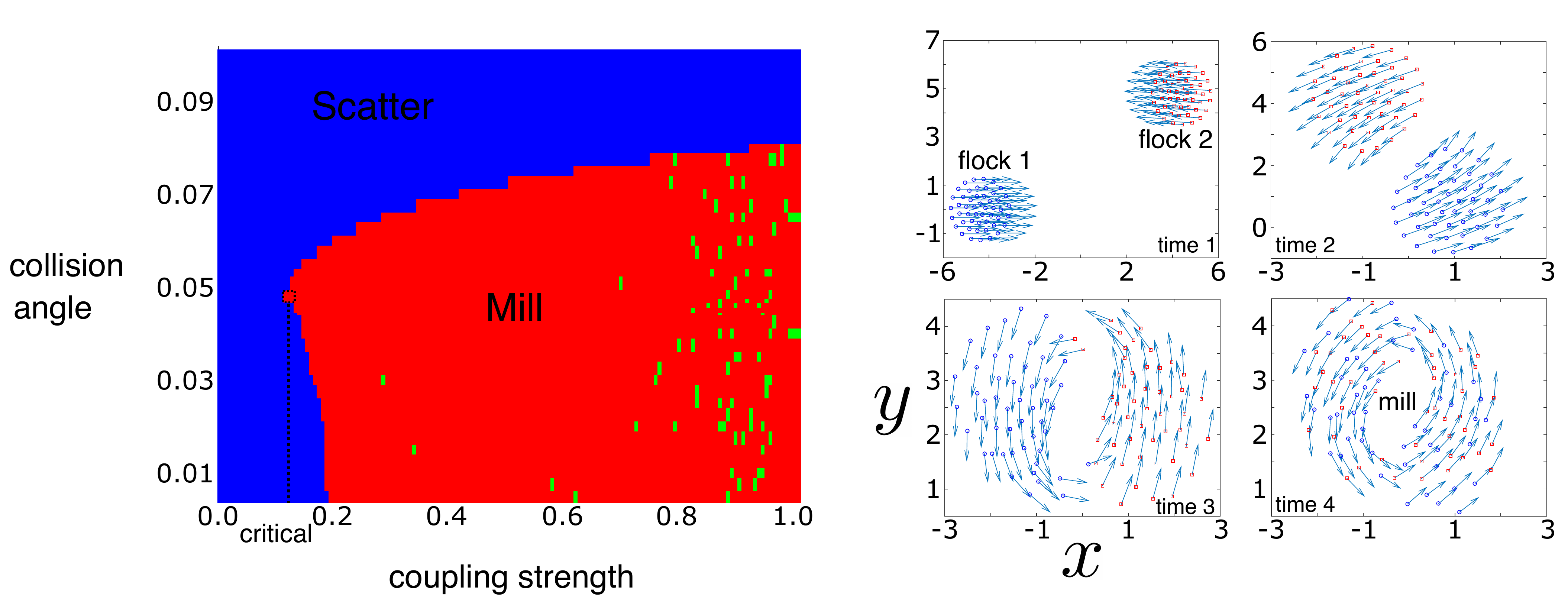}
\par\end{centering}
\caption{Two swarms colliding. A scattering diagram is shown on the
left that specifies the outcome of two-swarm collisions as a function
of the incidence angle and the coupling strength. On the right are
four time snapshots of the swarms at the critical point--
the minimum coupling, $\lambda_{min}$, at which a collision results in a
mill. Swarm parameters are $\alpha=1$, $\beta=5$, $C=10/9$, $l=0.75$, and $N=100$.}
\label{fig:MillingCollision}
\end{figure}

We now wish to concentrate on how one swarm may capture
another into a combined milling state where the combined center of mass is stationary
and the polarization is close to zero. To satisfy the latter, $\theta$ must be relatively small
so that the total momentum is near zero. We make a new diagram showing
exactly where the scattering-to-milling transition occurs for small $\theta$ as a function of coupling $\lambda$; 
an example is shown in Fig.~\ref{fig:MillingCollision}. The stable swarm states after collision are
specified with blue and red for scattering and milling, respectively; the
green portions indicate the formation of a combined flocking state, which is
comparatively infrequent for small $\theta$ (and decreases in frequency
as $N \rightarrow \infty$). 

In addition, in the right panel of Fig.~\ref{fig:MillingCollision}, 
we show an example of the approach to the milling
state as a series of time snapshots. Initially, the
swarms are far apart in flocking states with constant velocities.
As the two swarms approach, however, each agent begins to sense the forces of
intra-agent swarmers, causing the two swarms to    
rotate around each other while maintaining an approximately constant inter-swarm density. 
Over time the two swarms slowly relax to a well-mixed milling state composed
of uniformly distributed agents from both.

\begin{figure}
\begin{centering}
\includegraphics[scale=2]{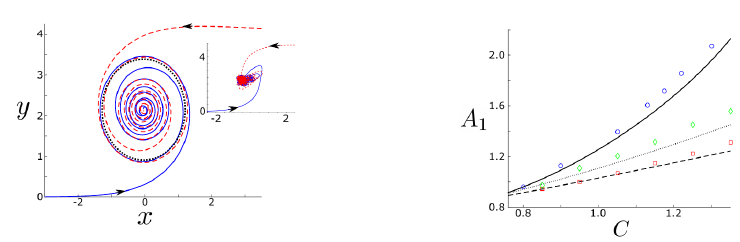}
\par\end{centering}
\caption{Collision dynamics resulting in milling. (a) Center-of-mass trajectories for two colliding swarms when $\lambda=\lambda_{min}$, shown with solid-blue and dashed-red lines. Arrows give the direction of motion. The dashed-black line indicates the bifurcating limit cycle in the uniform constant density approximation. Other swarm parameters are $\alpha=1$, $\beta=5$, $l=0.75$, $N=100$, and $C\!=\!1.0$. The inlet panel shows the corresponding trajectory for $\lambda=2 \lambda_{min}$. (b) Maximum x-coordinate reached by the center of mass of the rightward moving (blue) flock when $\lambda=\lambda_{min}$. Simulation results are shown with blue circles for $l=0.75$, green diamonds for $l=0.6$, and red squares for $l=0.5$. Limit-cycle predictions from theory are drawn with lines near each series. Other swarm parameters are $\alpha=1$, $\beta=5$, and $N=200$.}
\label{limitcycle}
\end{figure}

Motivated by Fig.~\ref{fig:MillingCollision}, one useful observation that can be made regarding the swarms is that, when flocking
towards a collision, each swarm behaves as a rigid body. Assuming such motion in the
swarms leads one to hypothesize that there exists a constant density
approximation when all agents have the same characteristics. Such an
approximation can used to create a theory for the center-of-mass dynamics
describing the approach to a milling state, as shown in
\cite{hindes2021critical}. In the left panel of Fig.~\ref{limitcycle}, the center of mass dynamics for
each swarm is shown at the critical coupling, $\lambda_{min}$: the smallest coupling, over all
collision angles, at which a milling state is stably formed.

The constant density theory predicts that in order for the milling state to occur, the dynamics
must approach a stable limit cycle of the interacting centers of mass. Within this approximation, 
the critical coupling corresponds to a generic saddle-node bifurcation. In general, the limit cycle acts as a capture radius, whereby the two
interacting flocks slowly converge to a common, stationary center
The same theory can be used to predict the maximum size of the transient center-of-mass oscillations as a function
of the repulsive coupling, $C$. In the right panel of Fig.~\ref{limitcycle} the
theory is plotted against numerical simulations to show how well the predictions
work for a range of different repulsive-force strengths.

\section{Analysis and final results of swarm  symmetry and asymmetry}

One interesting aspect of the theory is that it can provide a range of
parameter predictions for the critical coupling, $\lambda_{min}$, when the swarms are both symmetric and 
asymmetric. In particular, from the theory one can define the critical parameter for the saddle-node bifurcation via an
equation analogous to an escape-velocity relation, 
\begin{align}
\label{eq:escape}
v^{2}/2 -N\lambda_{min} V_{\text{eff}}(C,l)=0,
\end{align}
where $v$ is the speed of each flock, and $V_{\text{eff}}(C,l)$ quantifies the
strength of the potential between agents (see \cite{hindes2021critical} for
full mathematical details). In terms of scaling Eq.(\ref{eq:escape}) implies that, 
if the potential-forces and number of agents are held constant,
flocks moving twice as fast require four times the coupling in order to
capture. Similarly, flocks with twice as many particles must fly
$\sqrt{2}$-times faster in order to escape capture. 

\begin{figure}
\begin{centering}
\includegraphics[scale = 2]{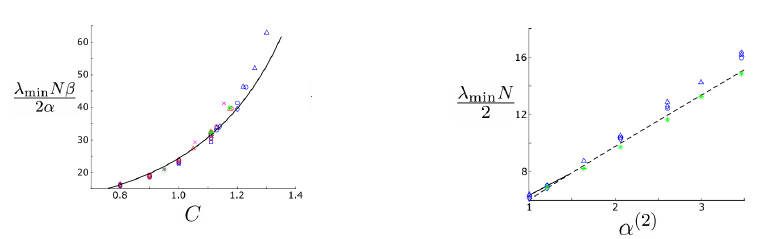}
\par\end{centering}
\caption{Critical coupling for forming milling states upon collision. (Left panel)
  Symmetric parameter collisions for $\alpha=1$ (blue) and $\alpha=2$
  (red): $N=10$ (squares), $N=20$ (diamonds), $N=40$ (circles),
  and $N=100$ (triangles). Green stars denote $\alpha=1$ and magenta
  x's denote $\alpha=2$, when 40 agents collide with 60. (Right panel) Asymmetric
  collisions for $C=10/9$ in which $\alpha^{(1)}=1$. Blue points
  indicate equal numbers in each flock: $N=20$ (diamonds), $N=40$
  (circles), and $N=100$ (triangles). Green stars denote collisions
  between 40 agents with $\alpha^{(1)}=1$ and $60$ agents with
  $\alpha^{(2)}$. Solid and dashed lines indicate theoretical predictions. Other swarm parameters are $\beta=5$ and
  $l=0.75$.}
\label{fig:sym_asym}
\end{figure}

We can use the theory to example how the velocity and potential function 
define the critical coupling $\lambda_{min}$ as we sweep physical swarm parameters. 
Examples are shown in Fig.~\ref{fig:sym_asym}. 
In the left subplot we show results for collisions with symmetric parameters.
Our predicted scaling collapse holds. Qualitatively, the critical coupling
increases monotonically with $C$, implying that the stronger the strength of
repulsion, the larger the coupling needs to be in order for colliding swarms
to form a mill. Also, note that our predictions are fairly robust to
heterogeneities in the numbers in each flock, particularly for smaller values
of $C/l-1$; predictions remain accurate for number asymmetries in the flocks
as large as $20\%$. In the right panel, we consider how theory compares in the asymmetric case of
two swarms with different velocities. In particular, agents in one flock have self-propulsion 
$\alpha_{i}=\alpha^{(1)}=1$, while $\alpha_{i}=\alpha^{(2)}$ is varied for the other flock.   
Again we see that when two swarms come together at the critical
coupling; the results between bifurcation theory and simulations agree well.
\\


\section{Preliminary colliding swarm experiments}
\begin{figure}[h]
  \centering
  \subfigure[Mixed Reality Setup]{
    \label{fig:mixed_reality}
    \includegraphics[width=0.65\linewidth]{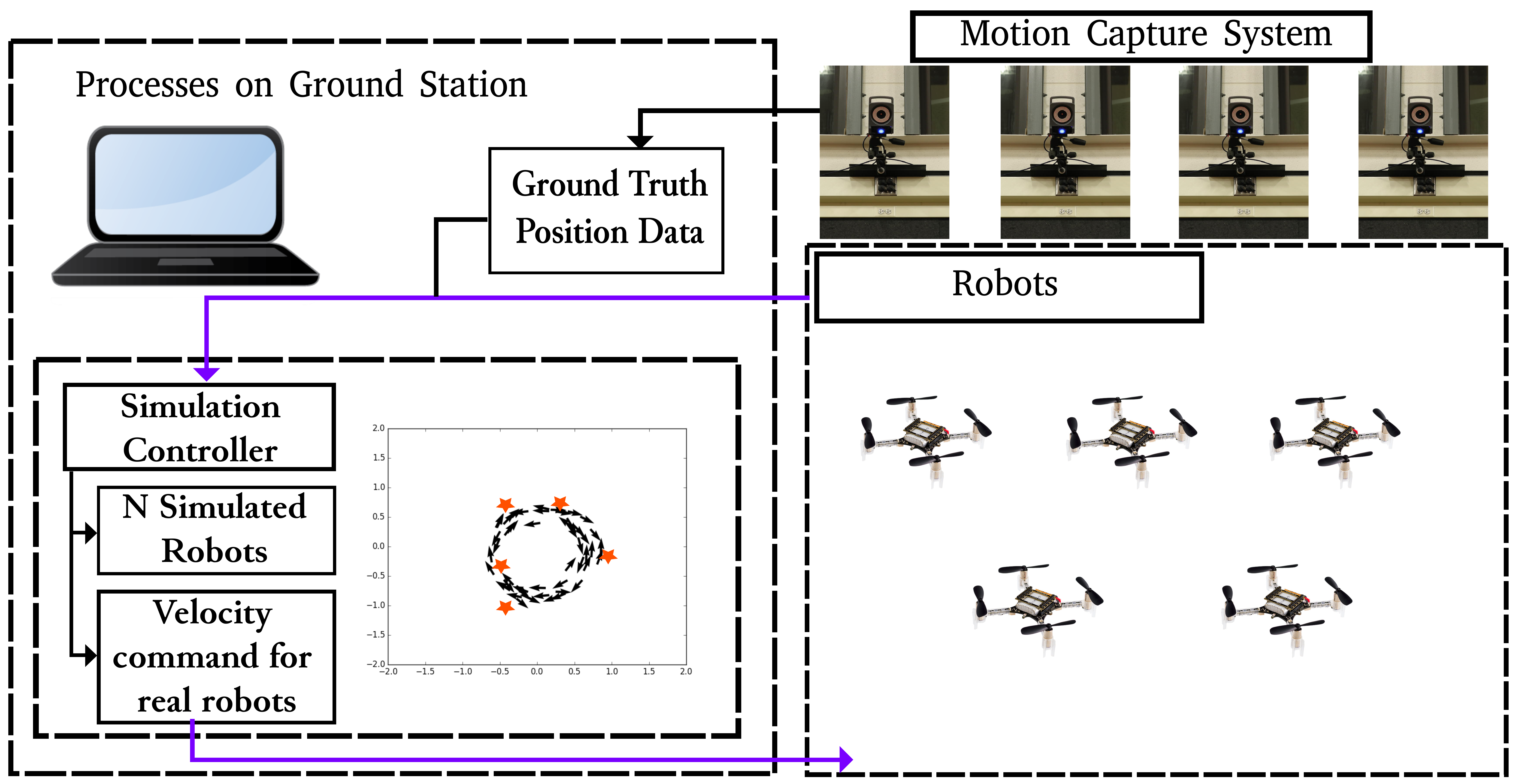}
    }
  \subfigure[Micro-UAV]{
    \label{fig:crazyflie}
    \includegraphics[width=0.23\linewidth]{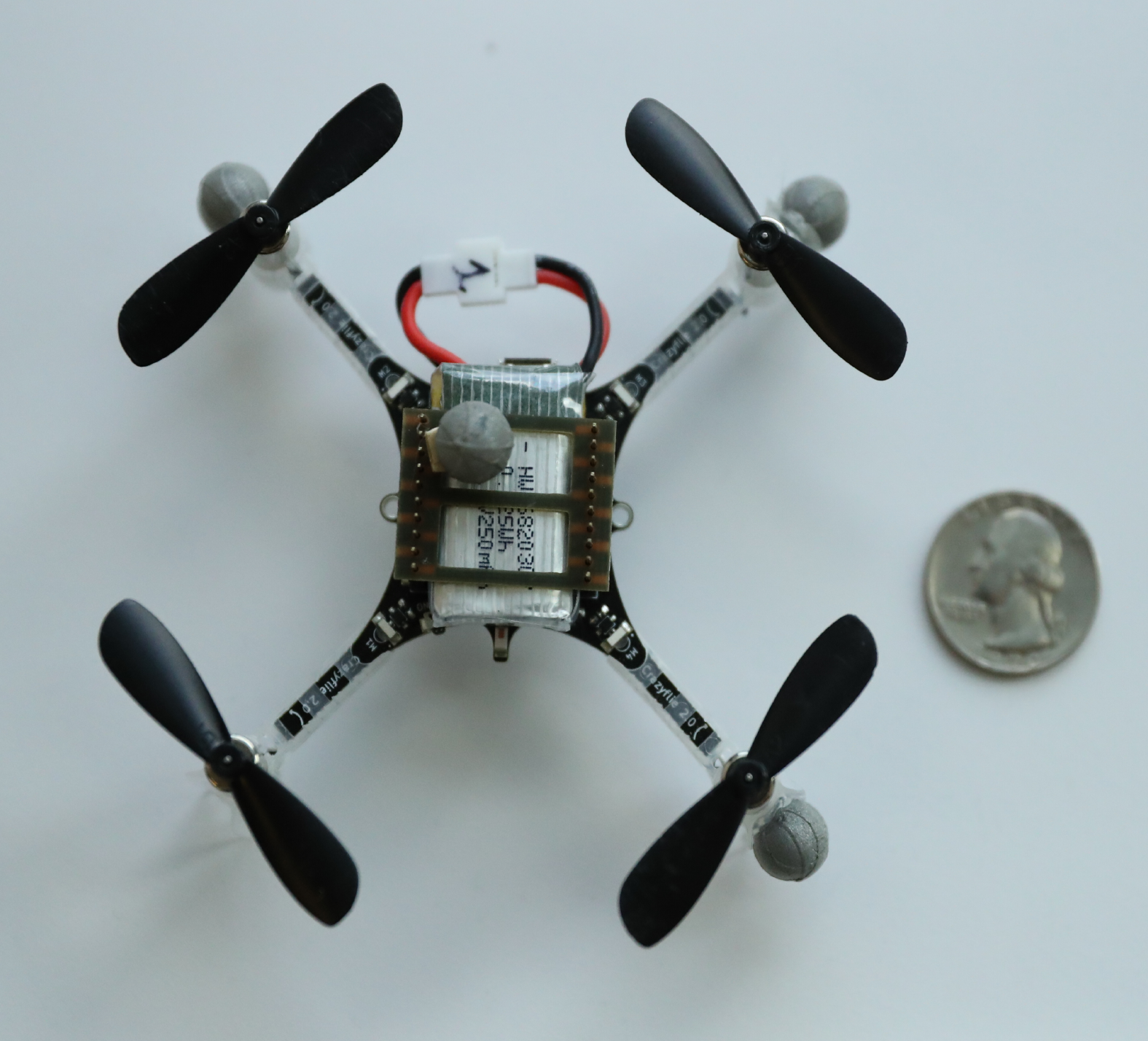}
  }
  \label{fig:experimental_setup}
  \caption{
    In Figure \ref{fig:mixed_reality} a mixed reality experimental platform is shown, which relies on each agent real and simulated having a global position and receiving some control command.
    In Figure \ref{fig:crazyflie} an example of the Crazyflie 2.0 Micro-UAV, which is used with the Crazyswarm Software. \cite{crazyswarm}.
  }
\end{figure}

We have begun to test our theoretical predictions in colliding swarm
experiments, where we implemented a mixed-reality setup\cite{szwaykowska2016collective,edwards2020delay}. To verify the
presented theoretical model we used up to eight Crazyflie micro-UAVs, shown in
Fig.~\ref{fig:crazyflie}; however eight is an insufficient number of robots to see
meaningful interaction between two large intersecting swarms. To help increase the
number of agents that were used during experimentation we used mixed reality
to couple real and virtual robots\cite{Szwaykowska2016}. In running the experiments, we used a dimensional
version of the Morse potential given by
\begin{align}
  U(r_i, r_j) = c_r e^{\frac{-|r_i - r_j|}{l_r}} - c_a e^{\frac{-|r_i - r_j|}{l_a}}.
  \label{eq:morse_potential}
\end{align}

\noindent where $c_r$ is the repulsion strength and $l_r$ is the length scale of the repulsion; likewise $c_a$ is the attraction strength and $l_a$ is the attraction length scale. 

\begin{figure*}
  \centering
  \subfigure[Eight real and eight virtual robots]{
    \label{fig:8v8}
    \includegraphics[width=0.85\linewidth]{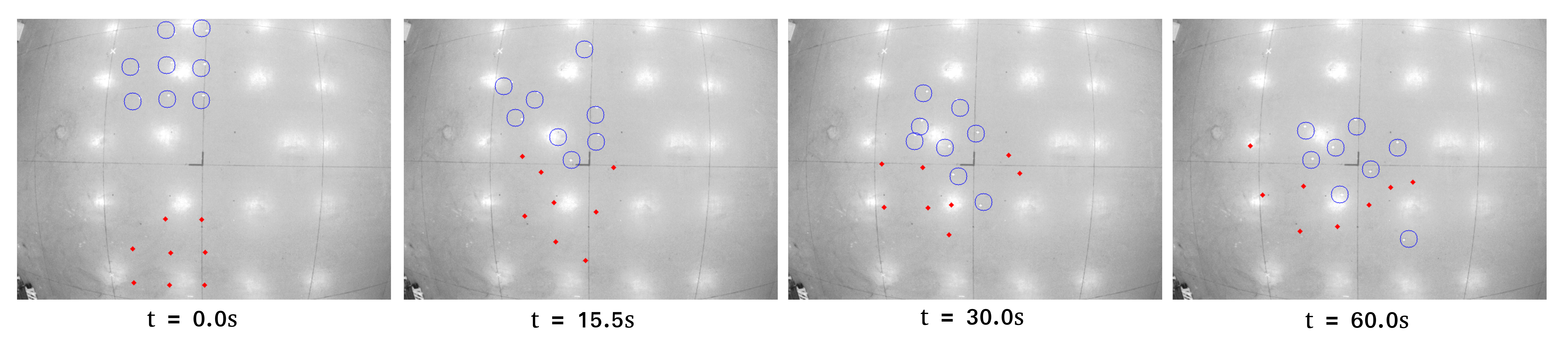}
    }
  \subfigure[High Repulsion]{
    \label{fig:5_real}
    \includegraphics[width=0.95\linewidth]{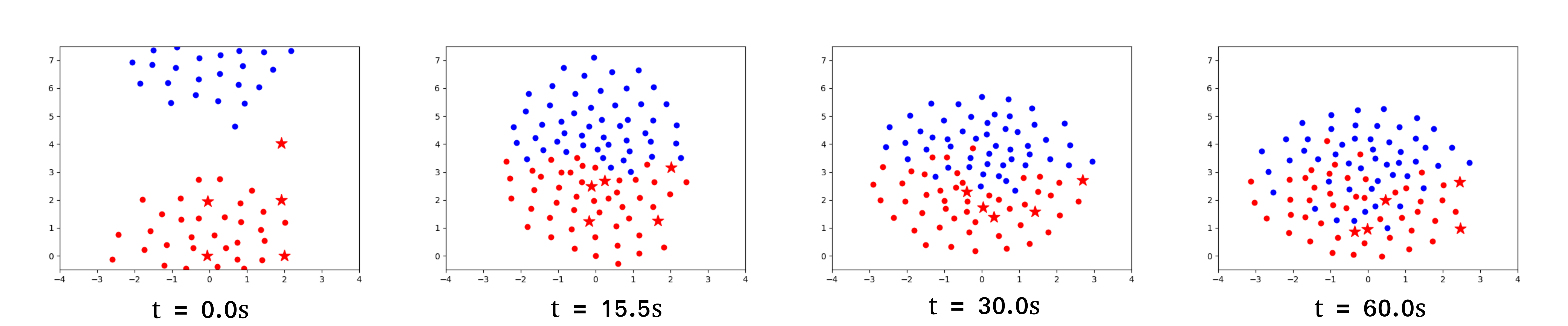}
  }
  \caption{
    An example of two time series mixed reality  experiments.
    Figure \ref{fig:8v8} shows 8 virtual colliding with 8 real robots. 
    Figure \ref{fig:5_real} shows an experiment in which 5 real robots
    join  with 45 simulated robots to collide with 50 simulated robots. }
  \label{fig:time_series}
\end{figure*}

The mixed reality system shown in Fig.~\ref{fig:mixed_reality} uses a Vicon
motion capture system in a 15x15m room with between 5-8 Crazyflie micro
UAV. The robots positions are shared through a ground station which also runs
the simulation. All agents positions are combined on the ground station and
new positions for the real robots are determined by using a double integrator
model of the agents. Figure \ref{fig:8v8} demonstrates how the physical robots
interact with simulated agents. The simulated agents, red dots, are projected
into the real world using a camera calibration and the real agents are
highlighted by blue circles.  These results allow for further improvement of
theoretical predictions and increase preparedness for field
experimentation.

Further examples of mixed reality experiments of two colliding swarms forming
a milling state with a stationary center of mass are shown in Figure
\ref{fig:time_series}. In addition to  eight real robots vs. eight simulated
robots and see a mill form, we consider even more agents where there are 5
real robots with 45 simulated robots versus 50 simulated robots. Due to the
inclusion of physical agents which require space between them it is necessary
to consider larger repulsion parameters, $c_r$ and $l_r$, to ensure robot safety. It is clear that even when these values are changed that experimentally a stationary mill is observed.

Although preliminary, the results show that when the theory is translated to
experiments, we can have one swarm capture another based on the physical
parameters chosen. Conversely, our theory and experiment should also
predict when colliding swarms will not form a milling state; i.e., based on
known parameters and sizes of the swarms, we can show one swarm cannot
capture another. Other measures beyond the polarization of how the colliding
swarms mix can also be ascertained;  one such  metric of measuring the scaling of the density of one swarm with
respect to another is presented in the Appendix A for the mixed reality experiments.

\section{Conclusion and Discussion}
Here we studied the collision of two swarms with nonlinear interactions, and focused in particular on predicting when 
such swarms would combine to form a mill. Unlike the full final-scattering diagram, which depends on whether or not 
a particular set of initial conditions falls within the high-dimensional basin-of-attraction for milling -- a hard problem in general, we concentrated 
on predicting the minimum coupling needed to sustain a mill after the collision of two flocks. By noticing that colliding swarms, which eventually form a mill, 
initially rotate around a common center with an approximately constant density, we were able to transform the question of a critical coupling into 
determining the stability of limit-cycle states within a rigid-body approximation. This approach produced predictions that were independent 
of initial conditions (only depending on physical swarm parameters) and provided a lower-bound on the critical coupling for
small collision angles. For example, in the case of symmetric flocks with equal numbers and physical parameters, the scatter-mill transition 
point was similar to an escape-velocity condition in which the critical
coupling scaled with the squared-speed of each flock, and inversely with the
number of agents in each flock. Our bifurcation analysis agreed well with
many-agent simulations.

Recent work in swarm robotics and autonomy has begun to address how one swarm can detect, redirect, capture, or 
defend itself against another\cite{8444217,9029573,9303837}. However, most approaches are algorithmic and lack 
basic physical and analytical insights. Our work fits nicely into the robotic swarm capture and redirect problem, since the critical coupling sets a general divide in parameter space between scattering and milling swarms operating with general physical interactions and dynamics. In this paper, however,
we have not included the effects of communication delays or
internal and external noise effects, which play a significant role
in swarms of mobile robots\cite{edwards2020delay,szwaykowska2016collective}.  
For example, it is known that when the
center of mass of a single swarm is stationary, time delays
in communication can result in stable oscillations in the center of mass itself. 
The oscillations are the result of a general delay-induced Hopf bifurcation. 
On the other hand, it is also known that (even) small amounts of noise
can act as a force, inducing large changes in swarm behavior\cite{6580546}.
Such large fluctuations may happen in the case where there are multiple attractors for the center of mass of
two interacting swarms. In such cases, noise ``kicks" 
the center of mass from one attractor to another. For these and other scenarios, new theory and
potential controls will have to be developed using some of the techniques we
have presented here to model how one flocking swarm can capture another.



\section{Appendix A}
In addition to polarization to quantify the nature of the milling state, it is
useful to find how one swarm embeds itself in another. One possible way to
achieve this is to  compute the local density of the swarm with respect to
another. To evaluate the level of interaction between the two swarms we
consider that for all agents in swarm $A$ we compute the following using swarm
$B$:

\begin{align}
  \gamma(r) = \sum_{i\in A}\sum_{j\in B} f(a_i, r, b_j),
  \label{eq:mixing}
\end{align}

\noindent where $f(a_i, r, b_j)$ is defined as follows:

\begin{equation}
  f(a_i, r, b_j) =
  \begin{cases}
    ||a_i - b_j|| < r & 1 \\
    ||a_i - b_j|| \geq r & 0,
  \end{cases} 
\end{equation}

\noindent where $r$ is the radius of inclusion.

  \begin{figure}[H]
  \centering
  
    \includegraphics[width=0.5\linewidth]{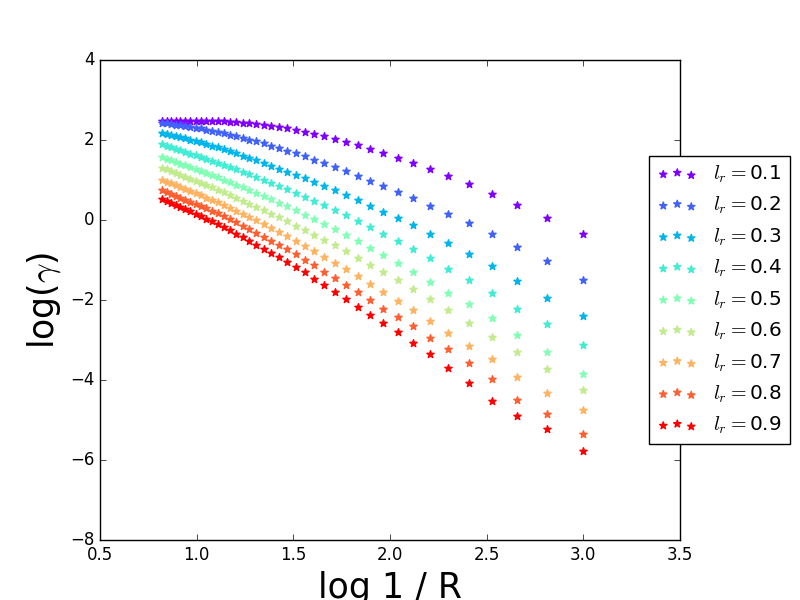}
    \label{fig:sim_results_swarm_density_strength}

  \caption{
    The plot shows how the $\gamma(R)$  changes as a function of ball size
    radius $R.$ The value of repulsion constant $c_r$ is varied  between $1.0
    - 5.5$ and fixed the repulsion length scale to be $l_r = 0.1$.}
  \label{fig:sim_results}
\end{figure}

This metric is calculating how many agents of a different swarm are in a local
neighborhood. It can be shown  $\gamma(r)$ is related
to the fractal or capacity dimension\cite{Mandelbrot}. By varying the 
radius of inclusion, one can see how the relative density varies, as shown in
Fig.~\ref{fig:sim_results}. Notice that there exists an inertial range where
$\gamma(R)$ exhibits roughly linear behavior, which signifies a scale
invariant local density of one swarm relative to another. We find that the mean
slope in the inertial region is $\mu = 1.59 \pm 0.19 $. The slope reflects a
dimension that is fractal, which implies that the agents of one swarm are
embedded in another in a complicated way. 


\end{document}